\documentclass{WileyMSP-template}
\usepackage{graphicx} 
\usepackage{subcaption}
\usepackage{xcolor}
\usepackage{soul}
\usepackage{float}
\usepackage{makecell}
\usepackage{amsmath}
\begin{document}

\pagestyle{fancy}
\rhead{\includegraphics[width=2.5cm]{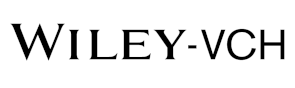}}

\title{Scalable Dip-Coated Bragg Mirrors for Strong Light–Matter Coupling with 2D Perovskites}

\maketitle

\newcommand{\hugo}[1]{{\color{blue}#1}}


\author{Melissa Méndez-Galván},
\author{Zaira Saucedo-Chávez},
\author{Diana Medrano},
\author{Michael Zuarez-Chamba},
\author{J Andrés Rojas-Sánchez},
\author{Yesenia A. García-Jomaso},
\author{César L. Ordóñez-Romero},
\author{Giuseppe Pirruccio},
\author{Arturo Camacho-Guardian},
\author{Galo J. A. A. Soler-Illia*} and
\author{Hugo A. Lara-García*}




\begin{affiliations}

M. Méndez-Galván, M. Zuarez-Chamba, G. Soler-Illia\\Instituto de Nanosistemas, Universidad Nacional de San Martín, Av. 25 de Mayo 1169, San Martín, Provincia de Buenos Aires, Argentina
Email Address: gsoler-illia@unsam.edu.ar

Z. Saucedo-Chávez, D. Medrano, J Andrés Rojas-Sánchez,Y. A. García-Jomaso, C. L. Ordóñez-Romero, G. Pirruccio, A. Camacho-Guardian, H. A. Lara-García\\
Instituto de Física, Depto. Física Química, Universidad Nacional Autónoma de México, Apartado Postal 20-364, 0100, Ciudad de México, México. 
Email Address: hugo.lara@fisica.unam.mx
\\

\end{affiliations}



\justify{
\begin{abstract}
{

We report a scalable and cost-effective method for fabricating high-performance Bragg mirrors using a bottom-up approach that combines evaporation-induced self-assembly (EISA) and dip-coating. The photonic crystals are composed of alternating mesoporous SiO$_2$ and dense TiO$_2$ layers, providing a high refractive index contrast ($\sim$0.8). This enables strong reflectance (up to 96\%) with as few as five bilayers and precise control of the photonic stop band across the visible spectrum by simply adjusting the deposition parameters.
Integration of a thin film of the two-dimensional perovskite (PEA)$_2$PbI$_4$ leads to strong light--matter coupling at room temperature. Angle-resolved reflectance and photoluminescence measurements reveal the formation of upper and lower polariton branches, with a Rabi splitting of 90 meV. The observed polaritonic dispersion is well described by a two-level system and Green’s function formalism.
This work demonstrates an efficient strategy for constructing tunable optical cavities using simple solution-based methods. The combination of high optical quality, spectral tunability, and strong coupling performance positions this platform as a promising candidate for low-threshold polariton lasers, nonlinear optics, and integrated optoelectronic devices.

}

\end{abstract}
}


\justify{
\section{Introduction}

One-dimensional photonic structures, such as Bragg mirrors, have attracted significant attention due to their ability to precisely control light transmission and reflection. This property makes them highly promising candidates for integration into advanced photonic devices. These structures consist of periodically alternating layers of materials with well-defined thicknesses and refractive indices, enabling the tuning of their optical properties, particularly the position and width of the photonic stop band, where maximum reflectivity is achieved.\cite{Faustini_1, ALMEIDA_2, ALMEIDA3, RABASTE4, Fuertes5} Optimizing the optical response and quality of photonic crystals requires highly precise fabrication of the multilayer systems, as even minor structural defects or inhomogeneities can lead to diffuse scattering. Consequently, achieving accurate thin film deposition is essential. In response to this need, significant efforts have been devoted to developing deposition techniques that offer stringent control over film thickness and morphology, such as sputtering and electron beam lithography. However, these methods typically require expensive equipment and stringent conditions, including ultra-high vacuum environments.\cite{Wang29} In recent years, the combination of Evaporation-Induced Self-Assembly (EISA) and dip-coating has emerged as a cost-effective, scalable, and versatile alternative for the fabrication of high-quality thin films. \cite{Soler31, Fuertes5,Hidalgo10, Grosso36} This approach enables fine control over layer thickness through thermal treatments, dehydration, and sintering processes, while maintaining compatibility with a wide range of materials. As such, it is particularly well-suited for producing advanced photonic structures.\cite{Colodrero6}
\\\\
One of the key advantages of using EISA and dip-coating synthesis methods is the ability to fabricate mesoporous films, where precise control over pore volume and diameter can be achieved by incorporating different templates during synthesis.\cite{Fuertes5, Hidalgo10,Innocenzi37} This control enables the tuning of the film’s optical properties, such as the refractive index. By adjusting the refractive index, it becomes possible to increase the contrast between the materials used in Bragg mirror construction,\cite{Fuertes5, Gazoni13, Calvo14} thus enabling the fabrication of efficient, high-reflectivity Bragg mirrors with a reduced number of bi-layers, which simplifies processing and enhances scalability.
\\\\
One-dimensional photonic crystals composed of sol–gel mesoporous and dense metal oxide layers have been explored for structures exhibiting responsivity to vapors, \cite{Fuertes5} plasmon-photonic crystal coupling \cite{Gazoni13} or graded optical responses,\cite{Faustini_1}. 
Mesoporous SiO$_2$ and dense TiO$_2$ layers with spatially varying properties have been shown to support Fabry-Perot resonance. However, the realization of high-contrast Bragg mirrors using this material pair remains an open area of research. Such systems are especially appealing for use in optical cavities tailored for strong light–matter coupling, where high reflectivity, low-loss interfaces, and refractive index tunability are essential.
\\\\
Beyond their structural and material considerations, Bragg mirrors are of particular interest due to their ability to form optical cavities that enable strong light–matter coupling between confined photonic modes and excitonic transitions. This regime gives rise to hybrid light–matter quasiparticles known as exciton–polaritons,\cite{Hopfield1958,Francisco16, Bujalance17,Bai2025}  which inherit the low effective mass and coherence of photons along with the strong nonlinearity of excitons~\cite{Polimeno2025}. These properties enable the observation of collective quantum effects such as polariton condensation,~\cite{deng2002condensation,Kasprzak2006,Deng2010} superfluidity~\cite{Amo2009}, and the formation of optical vortices.\cite{Lagoudakis2008,Strang18} As a result, exciton–polariton systems are increasingly explored for applications in nonlinear optics, quantum light sources, and low-threshold polariton lasers, paving the way for advanced optoelectronic and photonic technologies.
\\\\
In this context, the strong light–matter coupling between metal halide perovskites (MHPs) and optical resonators has garnered considerable attention, owning to the exceptional optoelectronic properties of MHPs. \cite{Moon24, Zhang23,Zhang21,Das22, Wang38} In particular, two-dimensional metal halide perovskites (2D-MHPs) incorporate bulky organic cations into the ABX$_3$ perovskite lattice, forming naturally self-assembled quantum well structures.\cite{Blancon27} These materials exhibit strong exciton confinement at room temperature. \cite{Lutong20, Ghosh26,Zhang23,Gomez28} Owing to these attributes, 2D-MHPs have emerged as highly attractive candidates for the investigation of strong light–matter coupling phenomena.\cite{Zhang23} For instance, Gomez-Dominguez et al.,\cite{Gomez28} demonstrated strong exciton–photon coupling in microcavities composed of the 2D perovskite (PEA)$_2$PbI$_4$ and a commercial distributed Bragg reflector (DBR) consisting of 21 dielectric bilayers. The microcavities were carefully designed by tuning the detuning parameter to enable optimal spectral overlap between the lower polariton branch and the exciton reservoir emission.  Despite these advances, performance remains limited by the presence of multiple radiative recombination pathways within the exciton reservoir, which reduce the efficiency of radiative pumping into the polariton ground state. Similarly, Wang et al.,\cite{Wang29} reported strong exciton–photon coupling in microcavities incorporating a 20.5-bilayer Bragg mirror fabricated via electron beam evaporation, combined with exfoliated 2D organic–inorganic perovskites, specifically (PEA)$_2$PbI$_4$ and (PEA)$_2$PbBr$_4$. In their study, coherent coupling between excitons, cavity modes, and Bragg modes produced clear anticrossing features and mode splitting, indicative of hybridized polariton states. Notably, they demonstrated that the polariton dispersion could be finely tuned by adjusting the thickness of the perovskite layers, thus enabling precise control over the photonic properties of the system.
These findings underscore the potential of 2D organic–inorganic perovskites as a powerful platform for studying strong light–matter coupling phenomena at room temperature, highlighting their relevance for practical polaritonic devices.
\\\\
In this work, we demonstrate that high-quality Bragg mirrors can be fabricated using a simple, cost-effective, accessible, and scalable bottom-up approach that combines Evaporation-Induced Self-Assembly (EISA) with dip-coating techniques. These mirrors are composed of alternating mesoporous SiO$_2$ and dense TiO$_2$ layers, which provide a high refractive index contrast (n $\approx$ 1.3 and n $\approx$ 2.1, respectively). With only five bilayers, this method yields reflectance values higher than 94 \%. Furthermore, the stop band of the resulting one-dimensional photonic crystals can be finely tuned across the entire visible spectrum by adjusting the withdrawal speed (which, in turn, controls the film thickness) during deposition. Notably, strong light–matter coupling at room temperature was achieved using the 2D perovskite (PEA)$_2$PbI$_4$, leading to the clear formation of upper and lower polariton branches. These results highlight the robustness and versatility of the proposed Bragg mirror architecture, paving the way for its integration into next-generation polaritonic platforms and tunable optoelectronic systems.

\section{Results and Discussion}
\subsection{Bragg Mirrors}

To understand the performance of the fabricated Bragg mirrors, it is essential to first characterize the structural and optical properties of the constituent layers, specifically the mesoporous SiO$_2$ and dense TiO$_2$ monolayers. These two materials form the periodically alternating architecture that defines the photonic stop band and governs the reflectance behavior of the system.
\\\\
The morphology of both types of monolayers was investigated by scanning electron microscopy (SEM). As shown in Figure S1a, the mesoporous SiO$_2$ film exhibits a homogeneously distributed porous network. However, partially blocked pores are also observed, which can be attributed to the incomplete thermal decomposition of the structure-directing agent Pluronic F127. Complete removal and carbonization of Pluronic F127 typically occur above 300 °C, whereas the films in this work were treated at 200 °C. Interestingly, this residual component contributes positively by preventing infiltration of precursor solutions during subsequent layer deposition, thereby preserving the integrity of the mesoporous architecture. In addition, we also studied the refractive index of fully calcined monolayers and found that it remains similar to that of films where the Pluronic F127 was not completely removed. On the other hand, Figure S1b presents the morphology of the dense TiO$_2$ monolayer, which shows a continuous and defect-free surface. The uniformity and stability of these films confirm that the synthesis protocol yields mechanically robust layers suitable for multilayer deposition. 
\\\\
The optical properties of the individual monolayers were characterized using spectroscopic ellipsometry. The mesoporous SiO$_2$ films exhibited a refractive index of approximately 1.3 at 623 nm, consistent with the presence of mesoporosity. In contrast, the dense TiO$_2$ layers exhibited a much higher refractive index of approximately 2.1 at the same wavelength, with negligible absorption in the visible range (Figure S2). This high refractive index contrast of nearly 0.8 is key to achieving a strong modulation of the dielectric function within the multilayer structure, which in turn leads to a well-defined stop band with high reflectivity, even when using a small number of bilayers.
\\\\
The Bragg mirrors were fabricated by alternately depositing mesoporous SiO$_2$ and dense TiO$_2$ layers via successive dip-coating. 
The photonic crystals are designated as ST-X-MB, where ST represents the SiO$_2$ and TiO$_2$ layers, X: 1, 2, 3 and 4 indicates the withdrawal speed used during dip-coating (with 1 corresponding to the lowest velocity and 4 to the highest) and MB (5B, 6B, or 7B) denotes the number of bilayers: five, six, or seven, respectively.
\\\\
The morphology of the ST-X-5B samples was analyzed by scanning electron microscopy (SEM), as shown in Figure 1. Cross-sectional micrographs reveal the periodic architecture of the Bragg mirrors, where the alternating mesoporous SiO$_2$ and dense TiO$_2$ layers are clearly distinguished. The TiO$_2$ layers appear brighter due to their higher electron density, whereas the SiO$_2$ layers show reduced brightness, attributed to their lower electron density resulting from a more porous structure. The thickness of each layer was measured using ImageJ, confirming a high degree of uniformity across the entire structure.The average layer thicknesses and the corresponding withdrawal speeds used during dip-coating are
summarized in Table 1.
For the mesoporous SiO$_2$ layers, the thickness ranged from a minimum of 101.0 ± 6.2 nm (ST-1-5B) to a maximum of 150.0 ± 12.2 nm (ST-4-5B), while for the dense TiO$_2$ layers, the values ranged from 39.6 ± 7.2 nm to 60.7 ± 8.6 nm for the same samples. These values reflect the precise tunability of film thickness by adjusting the withdrawal speed, in agreement with previous reports.\cite{Hidalgo10,Soler31} Additionally, in multilayered samples with a greater number of bilayers (ST-4-6B and ST-4-7B), SEM analysis (Figure S3) confirmed that the thickness of individual layers remained consistent throughout the stack, even after thermal treatment. However, a notable increase in structural fragility was observed as the number of bilayers increased, likely due to accumulated internal mechanical stress within the multilayer assembly.
\begin{table}[h]
\centering
\begin{tabular}{lcccc}
\hline
\textbf{Sample} & \makecell{\textbf{SiO$_2$}\\\textbf{thickness}\\(nm)} 
& \makecell{\textbf{Withdrawal}\\\textbf{speed}\\(mm s$^{-1}$)} 
& \makecell{\textbf{TiO$_2$}\\\textbf{thickness}\\(nm)} 
& \makecell{\textbf{Withdrawal}\\\textbf{speed}\\(mm s$^{-1}$)} \\
\hline
ST-1-5B & 101.0 ± 6.2 & 0.7 & 39.6 ± 7.2 & 0.6 \\
ST-2-5B & 112.7 ± 8.5 & 0.8 & 51.9 ± 5.0 & 0.7 \\
ST-3-5B & 142.8 ± 11.4 & 1.0 & 56.2 ± 9.5 & 0.9 \\
ST-4-5B & 150.0 ± 12.2 & 1.2 & 60.7 ± 8.6 & 1.1 \\
\hline
\end{tabular}
\caption{Thickness and withdrawal speed of SiO$_2$ and TiO$_2$ films.}
\label{tab:thickness}
\end{table}

\begin{figure*}[htbp]
   \centering
  \includegraphics[width=1\linewidth]{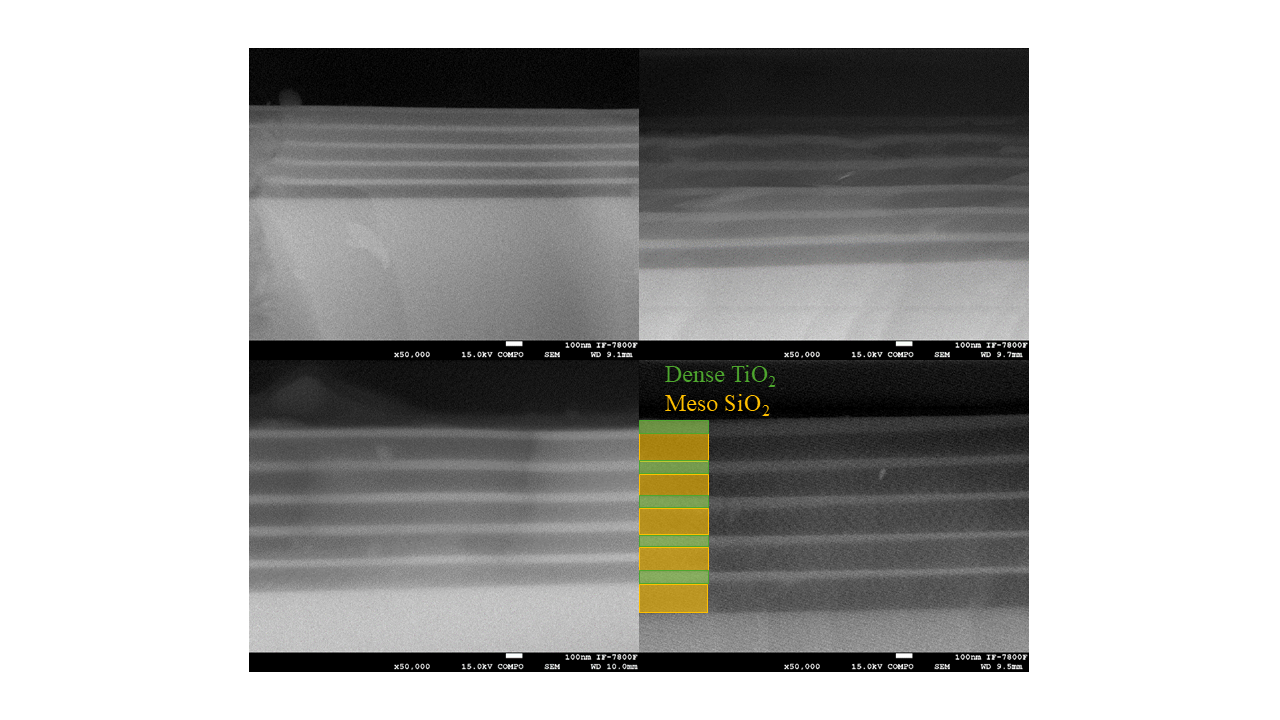}
  \caption{Scanning electron microscopy cross section images of the ST-1-5B, ST-2-5B, ST-3-5B and ST-4-5B samples.}
  \label{fig:sem} 
\end{figure*}
The optical properties of the Bragg structures and optical cavities were characterized by UV-Vis spectroscopy (Figures 2, 4, and S4).
Figure 2 shows the reflectance spectra of samples with five bilayers. The ST-1-5B sample exhibits a photonic stop band spanning 415–550 nm (centre wavelength 482 nm) with maximum reflectance of 96\%. The ST-2-5B structure’s stop band extends from 450–615 nm (centre 532 nm) with a peak reflectance of 96\%. For ST-3-5B, the band ranges from 525–700 nm (centre 612 nm) with 94\% reflectance, and ST-4-5B shows a stop band from 575–770 nm (centre 672 nm) with 94\% reflectance. The width of the stop band increases with layer thickness, which enhances the optical path difference between the alternating layers. This broadening is attributed to the refractive index contrast between the mesoporous SiO$_2$ and dense TiO$_2$ layers, enabling strong reflection over a wider spectral range. By controlling the withdrawal speed and consequently the layer thickness, the photonic stop band can be finely tuned across the entire visible range, offering a versatile and scalable approach for designing multilayer mirrors with tailored optical properties (Figure S5). The slight decrease in reflectance for ST-3-5B and ST-4-5B is attributed to inhomogeneities likely originating from accumulated thermal stress and increased layer thickness, as confirmed by SEM observations. Notably, reflectance values exceeding 90\% were achieved with only five bilayers, surpassing or matching previous reports that often require nine or more layers for comparable performance.\cite{RABASTE4, Hidalgo10, Bertucci34, Brundieu33, Romanova35} 
\\\\
To thoroughly analyze the reflectance behavior of the Bragg mirrors, a calculation based on the transfer matrix method (TMM) was performed. A system comprising five bilayers was modeled using alternating mesostructured SiO$_2$ and dense TiO$_2$ layers, with refractive indices equal to n = 1.3 and n = 2.1, respectively, as previously determined by ellipsometry.
As shown in Figure 3, the photonic stop band can be tuned across a wide spectral range by adjusting the thickness of the alternating layers. In this study, the layer thicknesses used in the simulations were: 105.0 nm / 46.0 nm, 113.0 nm / 52.0 nm, 140.0 nm / 56.0 nm, and 150.0 nm / 60.0 nm for samples ST-1, ST-2, ST-3, and ST-4, respectively. These values closely match those measured from SEM cross-sections, indicating excellent agreement between the calculated and experimental response of the structures. Moreover, both datasets confirm that increasing the thickness of the bilayers results in a redshift of the stop band, consistent with the expected optical response of these periodic structure.
In addition to the spectral position, the width of the stop band also increases with refractive index contrast, as predicted by the TMM. The substantial contrast between SiO$_2$ and TiO$_2$ in this system not only enhances reflectance but also yields broad photonic bands, allowing efficient reflection over a wide wavelength range. For all calculated configurations, reflectance values reached up to 90\%, highlighting the strong constructive interference generated within the multilayer system. This behavior aligns well with experimental results, where high reflectance values above 90\% were achieved with only five bilayers. These findings demonstrate that combining materials with a high refractive index contrast enables the fabrication of efficient Bragg mirrors with minimal layer count, broad stop bands, and excellent optical performance.

\begin{figure} [H]
 \centering
\includegraphics[width=0.7\linewidth]{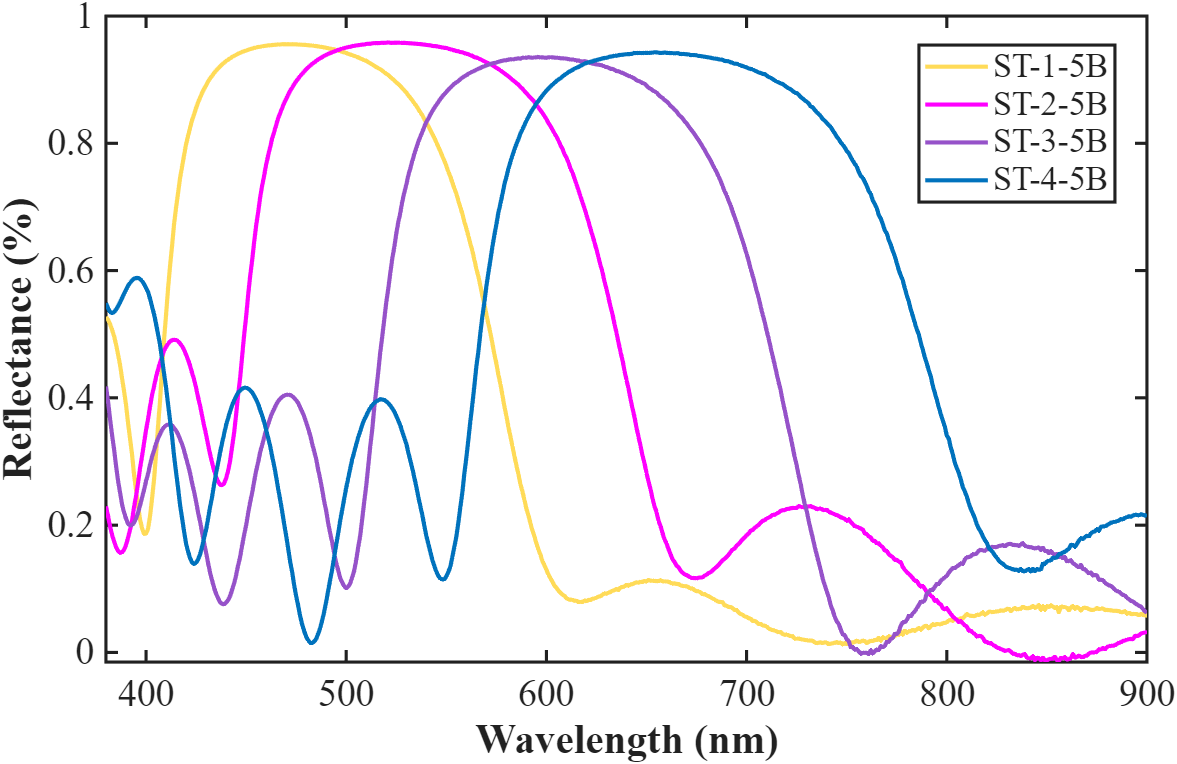}
\caption{Reflectance spectra of the ST-1-5B, ST-2-5B, ST-3-5B and ST-4-5B samples.}
\label{fig:boat1}   
\end{figure}

\begin{figure}[H]
\centering
\includegraphics[width=0.7\linewidth]{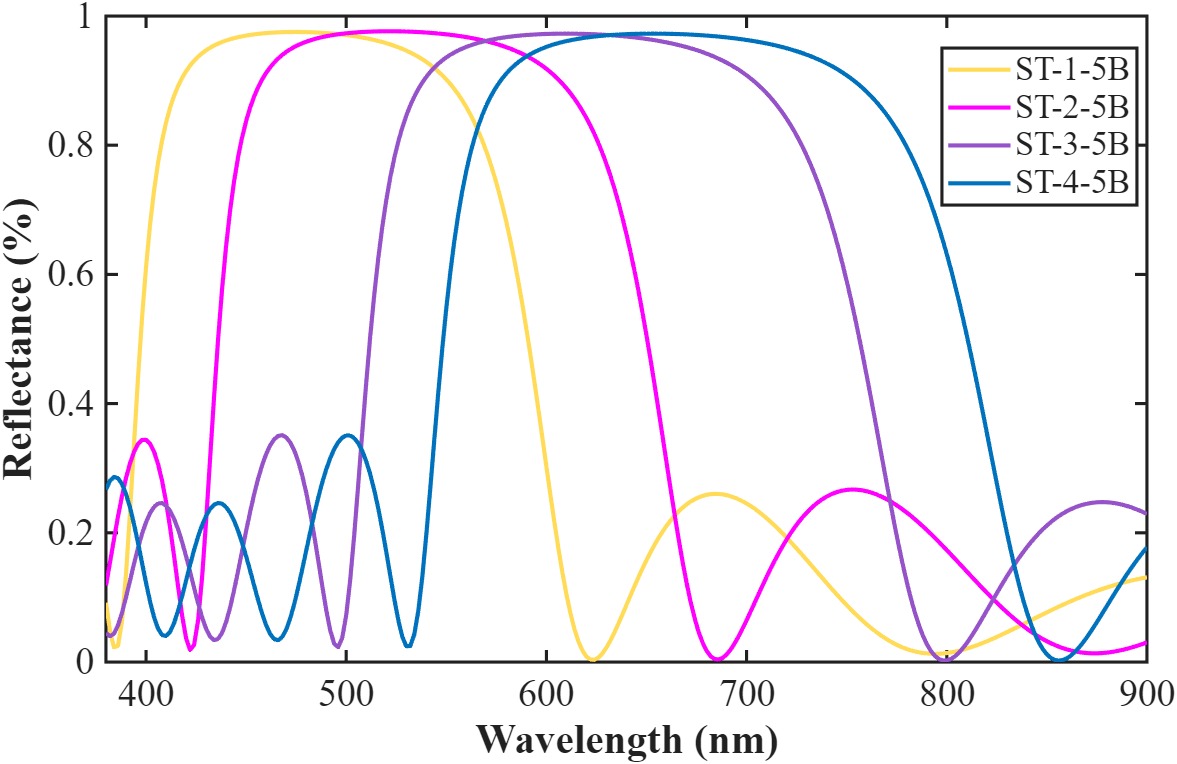}
\caption{Transfer matrix model for ST-1-5B, ST-2-5B, ST-3-5B and ST-4-5B samples.}
\label{fig:boat1}
\end{figure}

Furthermore, the influence of the number of bilayers on the optical response of the photonic crystals is illustrated in Figure 4. The reflectance spectra of samples ST-2-5B, ST-2-6B, and ST-2-7B reveal a systematic enhancement in peak reflectance, increasing from 96\% to 97\% and 98\% as the number of bilayers increases from five to seven. However, this improvement is accompanied by the emergence of oscillatory features at the high-wavelength edge of the stop band, which become more pronounced in the six- and seven-bilayer samples. These modulations are likely caused by structural inhomogeneities introduced during repeated annealing steps, leading to accumulated thermal stress and minor deviations in layer thickness or interface quality.
Although additional bilayers enhance overall reflectance, the appearance of these secondary interference fringes can degrade the spectral purity of optical cavities. Consequently, sample ST-2-5B was selected for cavity fabrication, as it offers a favorable trade-off between high reflectance and minimal spectral distortion.

\begin{figure}[H] 
\centering
  \includegraphics [width=0.7\linewidth] {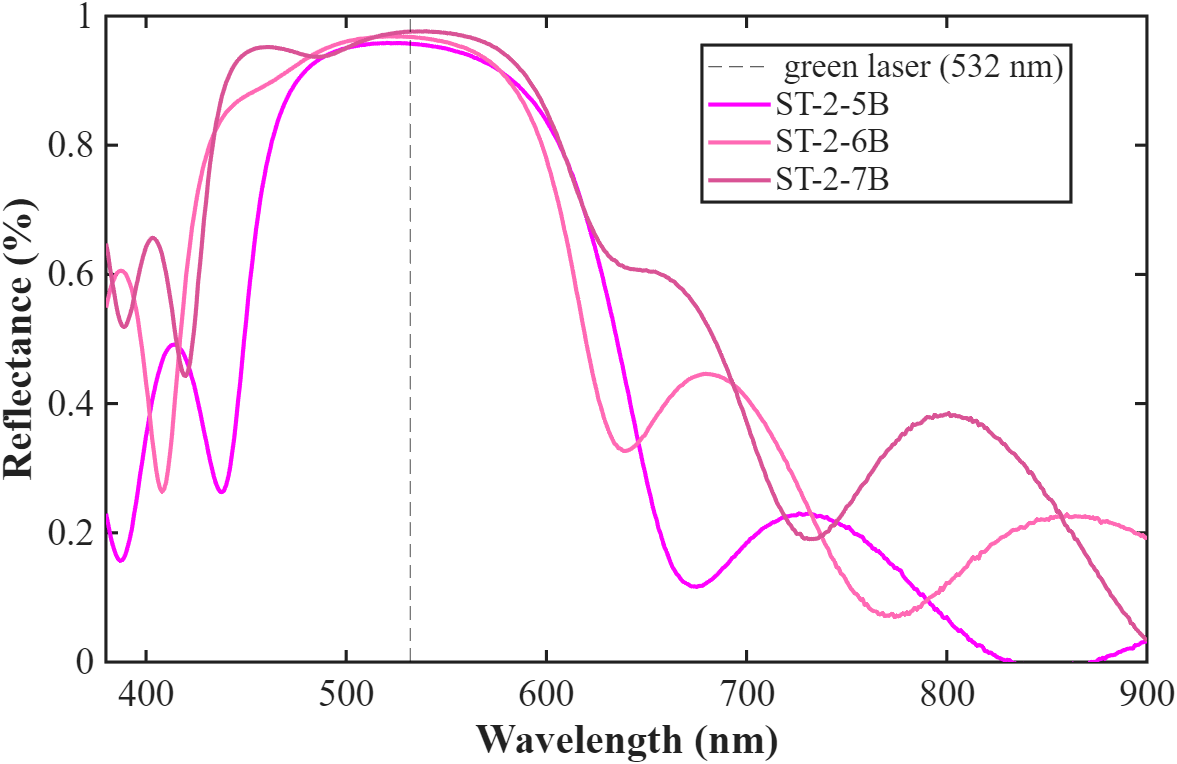}
  \caption{Reflectance spectra of the ST-2 sample for five, six and seven bilayers.}
  \label{fig:boat1}
\end{figure}

\subsection{2D-Perovskite}
Thin films of the two-dimensional perovskite (PEA)$_2$PbI$_4$ were deposited onto the Bragg mirrors via a solution-based spin-coating method, using a 2:1 molar ratio of phenylethylammonium iodide ((PEA)I) and lead(II) iodide (PbI$_2$), dissolved in \textit{N,N}-dimethylformamide (DMF). Among the tested precursor concentrations (3 M, 1.3 M, and 0.23 M), the 0.23 M solution yielded films with the lowest surface roughness , high crystallinity, and a uniform thickness of approximately 80 nm (Figure S6).
To evaluate the surface topography and roughness of the 0.23 M (PEA)$_2$PbI$_4$ film, atomic force microscopy (AFM) analysis was performed (Figure~S7). The scanned area (25 $\mu$m~$\times$~25~$\mu$m) revealed grain sizes ranging from 1 to 5 $\mu$m and a surface roughness below 10 nm, in agreement with the SEM micrographs. These results confirm that a 0.23~M precursor concentration enables the formation of uniform films with low roughness. This optimized condition was therefore selected for all subsequent optical characterization and microcavity integration.
\\\\
To verify the crystallographic structure of the perovskite, X-ray diffraction (XRD) analysis was performed (Figure~5). The resulting diffractogram displays sharp and intense peaks corresponding to the characteristic planes of the (PEA)$_2$PbI$_4$ phase,\cite{Gomez28,Lutong20} with no evidence of secondary phases, confirming the phase purity and structural integrity of the deposited film.
\\\\
The optical properties of the optimized (PEA)$_2$PbI$_4$ layer were evaluated through UV--Vis absorption and photoluminescence (PL) spectroscopy (Figure 6). A strong and narrow excitonic absorption peak is observed at 517 nm, while the PL spectrum exhibits a well-defined emission maximum at 528 nm, resulting in a small Stokes shift of approximately 11 nm. This narrow shift suggests low exciton--phonon coupling and minimal trap-assisted recombination, indicative of a high-quality film. The pronounced excitonic features and emission confined in the green spectral range confirm the material’s suitability for coupling with photonic modes within the visible stop band of the ST-2-5B Bragg mirror.
\\\\
In light of these properties, a hybrid Fabry--Pérot microcavity was constructed to explore light--matter interactions with the excitonic transition of (PEA)$_2$PbI$_4$. The microcavity consists of a bottom dielectric Bragg mirror, the (PEA)$_2$PbI$_4$ thin film as the active medium, and a semi-transparent 30 nm silver top mirror (Figure S9). To prevent parasitic charge transfer between the perovskite and the mirrors, thin interlayers of polymethyl methacrylate (PMMA) were spin-coated between each component. In addition, encapsulation with PMMA significantly improves the surface uniformity of the cavity structure by reducing topographical defects and suppressing morphological discontinuities, as confirmed by SEM analysis (Figure S8).
\\\\
The fabrication process involved the sequential spin-coating of PMMA onto a 10 $\times$~10~mm$^2$ Bragg mirror substrate, followed by thermal annealing at 60\,$^\circ$C for 5 minutes. The (PEA)$_2$PbI$_4$ film was then deposited from the 0.23 M precursor solution and annealed at 100\,$^\circ$C for 10 minutes to promote crystallisation. A final PMMA spacer layer was applied before sputtering a 30 nm Ag top mirror. This architecture enables efficient exciton--photon coupling within the optical cavity, laying the foundation for the exploration of strong coupling regimes in 2D lead halide perovskites.

\begin{figure} [H]
\centering
  \includegraphics[width=0.7\linewidth]{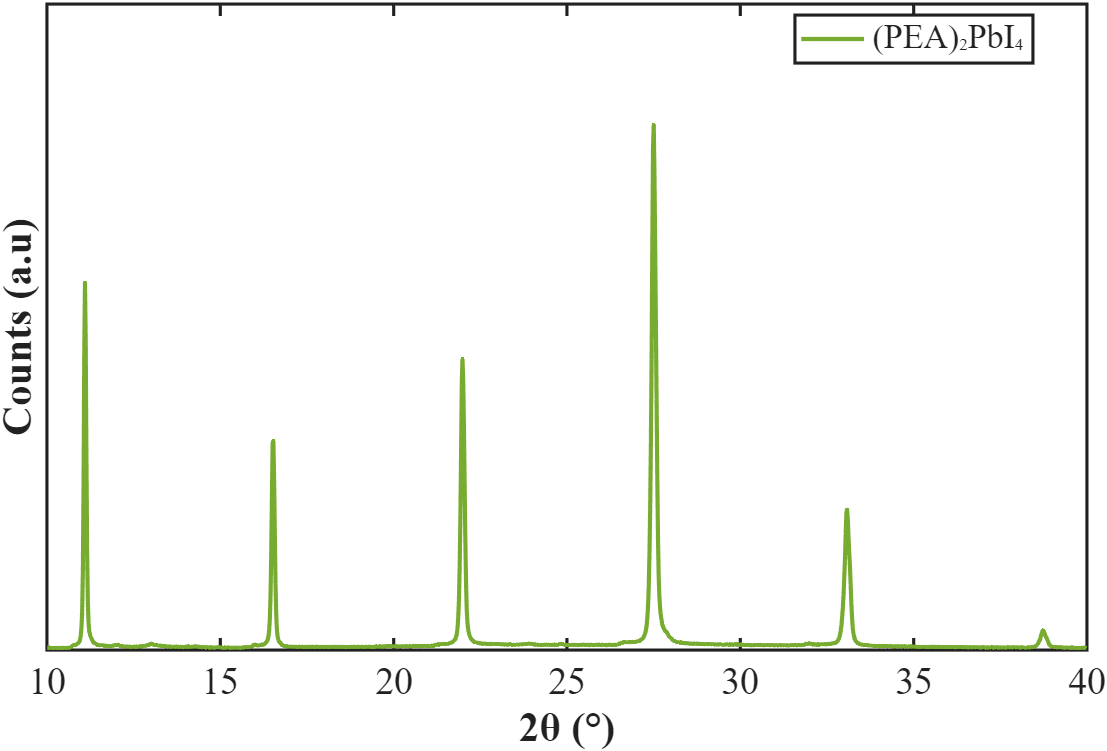}
  \caption{Difractogram of the (PEA)$_2$PbI$_4$ perovskite coupled within the  ST-2-5B sample.}
  \label{fig:boat1}
\end{figure}

\begin{figure} [H]
\centering
\includegraphics[width=0.7\linewidth]{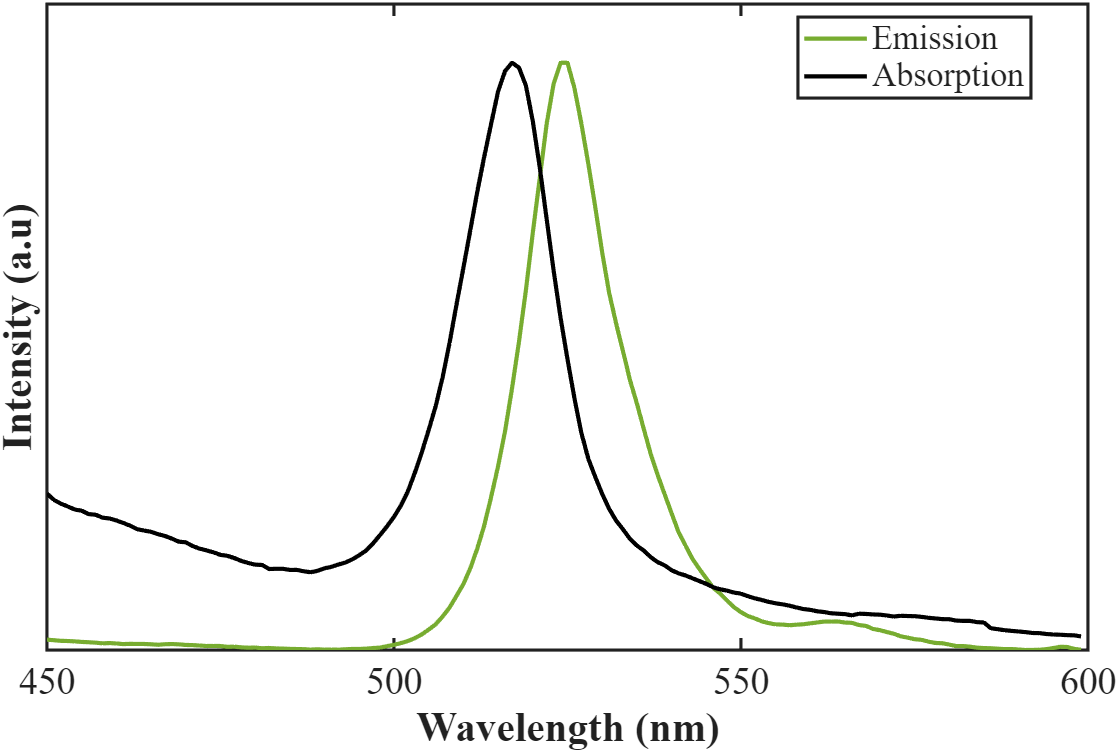}
 \caption{Absorption and emission spectra colected for the (PEA)$_2$PbI$_4$ perovskite.}
  \label{fig:boat1}
\end{figure}

\subsection{Polaritons: Experiment and Theory}

We now turn our attention to the experimental observation and theoretical modeling of polaritons in our system. Figures~\ref{fig:polaritons}(a)–(c) display angle-resolved reflectance spectra measured at different cavity detunings, defined as $\delta = \omega_c - \omega_X$, while panels~(d)–(f) show the corresponding theoretical calculations. Our fabrication method offers several key advantages: it enables the construction of optical cavities with precisely tunable photonic stop bands (from 425 to 770 nm), high reflectance (94–96\% for five bilayers), and excellent reproducibility. Most importantly, the optical quality and tunability of these resonators allow access to the strong light–matter coupling regime, as evidenced by the formation of well-resolved upper and lower polariton branches. Moreover, the ability to continuously tune the cavity detuning $\delta$ makes this platform ideal for exploring polariton physics across a broad range of coupling regimes which we now discuss in detail.
\\\\
To describe the polaritons, we consider a two-level Hamiltonian that captures the hybridization between cavity photons and excitons:
\begin{gather}
 \hat{H}(\theta) =
 \begin{bmatrix}
 \omega_c(\theta) & \Omega \\
 \Omega & \omega_X
 \end{bmatrix},
\end{gather}
where $\omega_c(\theta)$ denotes the angular-dependent cavity photon energy, $\omega_X$ is the exciton resonance, and $\Omega$ is the vacuum Rabi coupling. In the strong light–matter coupling regime, the eigenstates of this Hamiltonian correspond to hybrid quasiparticles known as polaritons. Their dispersion relations are given by:
\begin{gather}
\label{EqPol}
 E_{\text{UP}/\text{LP}}(\theta) = \omega_X + \frac{\delta \pm \sqrt{\delta^2 + 4\Omega^2}}{2},
\end{gather}
where $E_{\text{UP}}$ and $E_{\text{LP}}$ denote the energies of the upper and lower polariton branches, respectively. These theoretical dispersions are overlaid on the experimental data in Figs.~\ref{fig:polaritons}(a)–(c), with the upper (lower) polariton shown as a red (white) line. We find excellent agreement between theory and experiment.
\\\\
To quantitatively model the reflectance spectrum, we employ a Green’s function approach:
\begin{gather}
\label{Gfunction}
 \mathcal{G}(\theta, \omega) = \left[\mathcal{I}_{2\times2} \, \omega - \hat{H}(\theta) + i\Gamma \right]^{-1},
\end{gather}
where the damping matrix $\Gamma = \text{diag}(\gamma_c, \gamma_X)$ accounts for cavity photon losses ($\gamma_c$) and exciton linewidth broadening ($\gamma_X$). This Green’s function is directly connected to the experimentally measured reflectance via:
\begin{gather}
 R(\theta, \omega) = 1 + 2\gamma_c \, \text{Im}\left[ \mathcal{G}_{11}(\theta, \omega) \right] + \gamma_c^2 \left| \mathcal{G}_{11}(\theta, \omega) \right|^2.
\end{gather}
\\\\
For the theoretical modeling, we use a Rabi splitting of $2\Omega = 180\,\text{meV}$, exciton energy $\omega_X = 2.42\,\text{eV}$, and an effective refractive index $n_{\text{eff}} = 1.45$. The damping rates are chosen such that $2\Omega/\gamma_X = 12$ and $2\Omega/\gamma_c = 9$, ensuring that the system operates deep in the strong coupling regime.
\\\\
Figures~\ref{fig:polaritons}(a) and (d) correspond to zero detuning ($\delta / 2\Omega = -0.05$), where the upper and lower polariton branches exhibit symmetric anticrossing behavior. For positive detuning ($\delta / 2\Omega = 0.17$), shown in Figs.~\ref{fig:polaritons}(c) and (f), the polariton branches shift to higher energies: the upper polariton becomes more photonic and displays greater dispersion, while the lower polariton becomes more excitonic, resulting in a flatter dispersion due to its increased effective mass. As the detuning increases further, the polariton branches asymptotically approach the bare cavity and exciton energies, as clearly seen in the flattening of the lower polariton branch shown in  Figures~\ref{fig:polaritons}(c) and (f).
for $\delta / 2\Omega = 0.56$
\begin{figure}[H]
\centering
\includegraphics[width=0.8\linewidth]{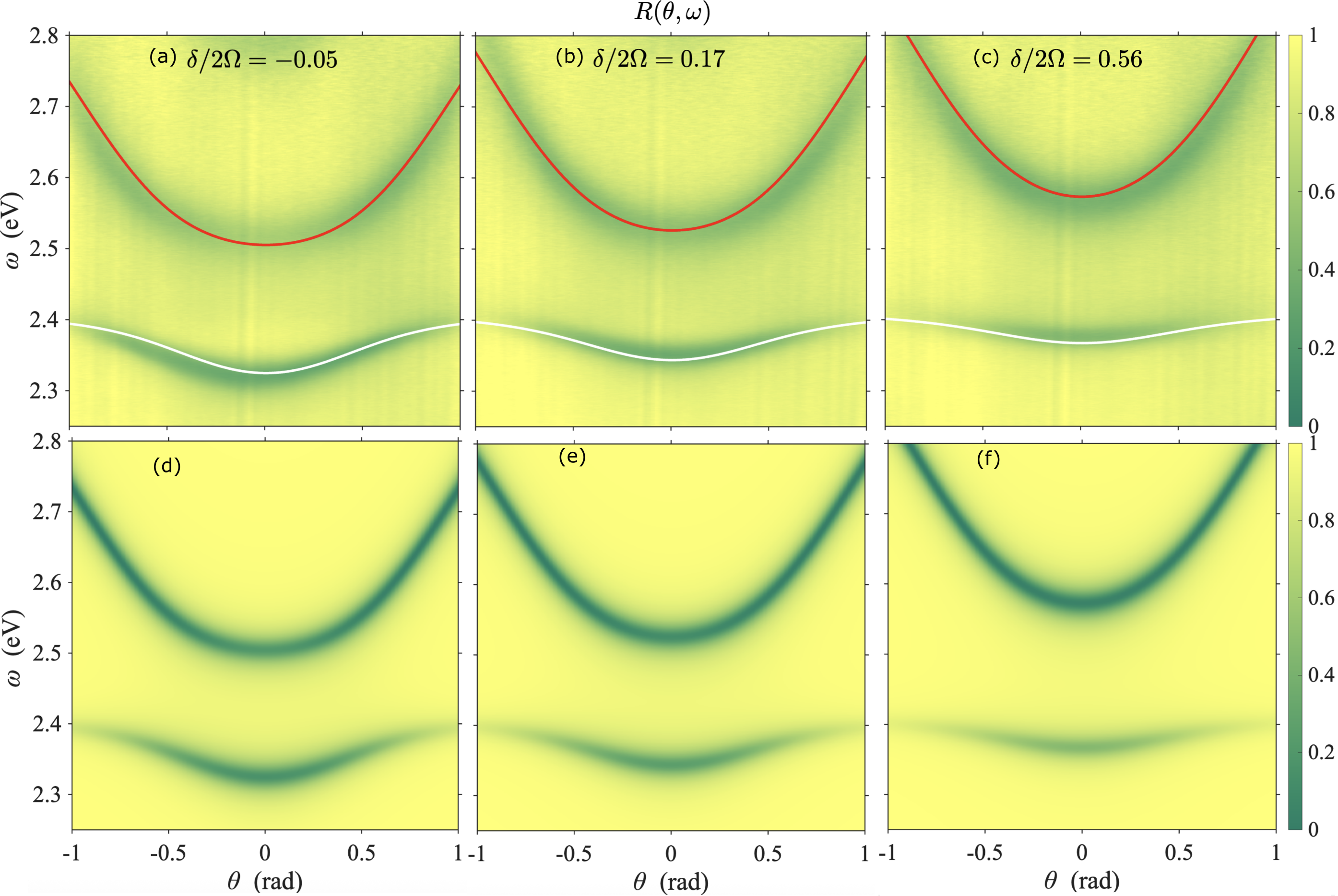}
\caption{\textbf{Experimental and theoretical reflectance spectra showing polariton formation.} 
Panels (a)–(c): Angle-resolved experimental reflectance for cavity-exciton detunings $\delta/2\Omega = -0.05$, $0.17$, and $0.56$, respectively. Panels (d)–(f): Corresponding theoretical reflectance spectra calculated via Green’s function formalism using the Hamiltonian in Eq.~(1). The red and white lines indicate the upper and lower polariton branches obtained from Eq.~\eqref{EqPol}, showing excellent agreement with experiment. As the detuning increases, the upper polariton becomes increasingly photonic, while the lower polariton becomes more excitonic and flatter in dispersion, highlighting the tunable hybrid nature of the polariton modes.}
\label{fig:polaritons}
\end{figure}

To further understand the hybrid nature of the polaritonic states, we performed angle-resolved photoluminescence (PL) spectroscopy under non-resonant excitation. The resulting emission spectra, shown in Fig.~\ref{fig:PLpolaritons}, display clear lower polariton branches for the same detuning values observed in reflectance measurements. The PL intensity follows the lower polariton dispersion with excellent agreement, as indicated by the overlaid theoretical curves obtained from Eq.~\eqref{EqPol}, confirming the strong coupling regime.
\\\\
For the smallest detuning, the emission is distributed at larger in-plane momenta along the lower polariton branch. The lack of significant emission near \( k_{\parallel} = 0 \) suggests inefficient relaxation toward the bottom of the branch, akin to the bottleneck effect.
\\\\
As the detuning increases, the emission is more uniformly distributed over the whole lower polariton band. We observe that polaritons predominantly populate energy regions of the dispersion where there is stronger spectral overlap of the lower polariton branch with the exciton emission. This indicates that the polariton emission is governed by radiative pumping \cite{Gomez28} with limited relaxation efficiency.
\\\\
Together with the reflectance data, these PL measurements provide strong evidence for the formation of exciton--photon hybrid states and reveal that polariton population dynamics in our system are controlled by detuning and the spectral properties of the exciton reservoir, rather than by thermal equilibrium processes.

\begin{figure}[H]
\centering
\includegraphics[width=1\linewidth]{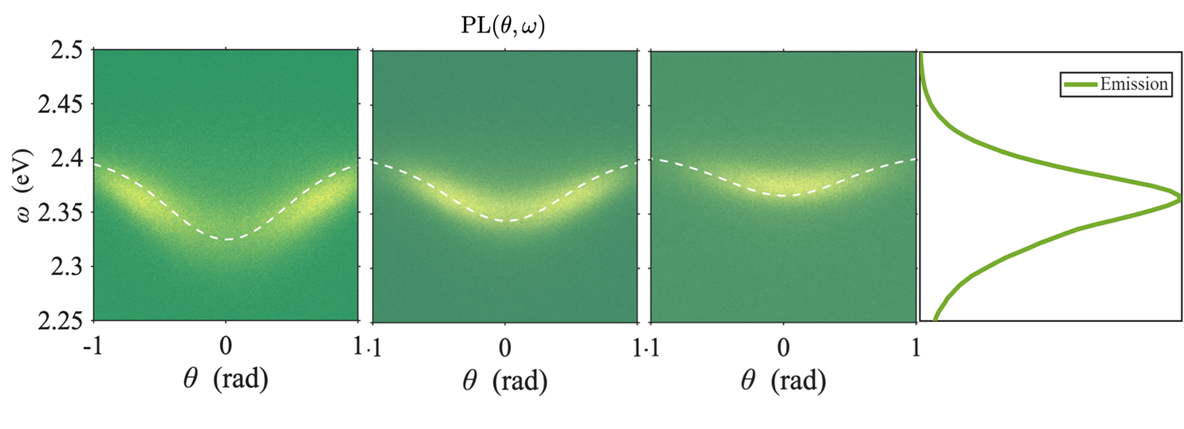}
\caption{\textbf{Angle-resolved photoluminescence spectra revealing polariton emission in arbitrary units.} 
Photoluminescence intensity as a function of emission angle $\theta$ and energy $\omega$ for three different cavity-exciton detunings: $\delta/2\Omega = -0.05$, $0.17$, and $0.56$ (left to right). The dashed white lines correspond to the lower polariton branch calculated from Eq.~\eqref{EqPol}, matching the peak emission across the dispersion. As the detuning increases, the polariton emission shifts and narrows, reflecting the increasing excitonic character of the lower polariton and its reduced radiative decay. These spectra complement the reflectance data in Fig.~\ref{fig:polaritons}, confirming the hybrid light-matter nature of the polaritonic modes.}
\label{fig:PLpolaritons}
\end{figure}




\section{Conclusion}

In summary, we have demonstrated a robust, scalable, and cost-effective method to fabricate high-quality Bragg mirrors via the combination of EISA and dip-coating techniques. By alternating mesoporous SiO$_2$ and dense TiO$_2$ layers, we achieved photonic crystals with high reflectance ($>$94\%) and finely tunable stop bands using as few as five bilayers significantly reducing the complexity compared to traditional DBR fabrication methods.
\\\\
When coupled with thin films of the 2D perovskite (PEA)$_2$PbI$_4$, these cavities exhibit clear evidence of strong exciton–photon coupling at room temperature, including well-resolved upper and lower polariton branches and a Rabi splitting of 90 meV. The optical response is in excellent agreement with theoretical models based on a two-level system and Green’s function formalism.
\\\\
Our approach opens a promising pathway for integrating solution-processed photonic components into next-generation polaritonic platforms. The simplicity and tunability of this system make it highly suitable for scalable photonic and optoelectronic devices, including low-threshold lasers, nonlinear optical elements, and quantum light sources.

}

\justify{
\section{Experimental Section}
\threesubsection{Bragg Mirrors}\\
Bragg mirrors were fabricated using a multilayered system composed of alternating mesoporous SiO$_2$ and dense TiO$_2$ films. These multilayers were prepared through a controlled dip-coating process followed by thermal stabilization. All reagents used were of analytical grade. Microscope glass slides (Marienfeld Superior™) were used as substrates and were sequentially cleaned with water, acetone, and finally ethanol.
\\\\
Mesoporous silica sol was prepared using tetraethyl orthosilicate (TEOS) as the inorganic precursor, absolute ethanol, and Pluronic F127 as a pore-direct agent. For this, two separate solutions were made: Solution A followed the molar ratio TEOS:C$_2$H$_6$O:HCl:H$_2$O; 1:3:0.00005:1 and was subjected to vigorous stirring at 23 °C for one hour; Solution B was prepared following the molar ratio C$_2$H$_6$O:HCl:H$_2$O:F127; 37:0.00795:9:0.005 and similarly maintained under vigorous stirring and recirculation at 23 °C for one hour. Subsequently, the solution B was carefully added to solution A. The resulting mixture was aged under constant stirring for 80 hours.
\\\\
Dense titania sol was synthesized using titanium isopropoxide (Ti(iOPr)$_4$) as the inorganic source, absolute ethanol, and acetylacetone (Acac,C$_5$H$_8$O$_2$) as chelating agent. The titania sol was prepared following the molar ratio Ti:C$_2$H$_6$O:HCl:Acac:H$_2$O; 1:40:4:1:15.
\\\\
Deposition of mesoporous SiO$_2$ films was performed via dip-coating at controlled withdrawal speed of 0.7, 0.8, 1.0, or 1.2 mm s$^{-1}$ under a relative humidity (RH) of 40 – 45\%. Post-deposition, films were aged in a 50\% RH chamber for 30 minutes, followed by thermal treatment at 60 °C and 130 °C (30 minutes each), and finally at 200 °C for one hour. The films were then cooled to ambient temperature.
\\\\
Dense titania films were deposited over the previously prepared mesoporous SiO$_2$ films at a constant withdrawal speed of 0.6 mm s$^{-1}$, 0.7 mm s$^{-1}$, 0.9 mm s$^{-1}$, or 1.1 mm s$^{-1}$ under controlled humidity (RH) of 20 \%. The films were stabilized at 200 °C for 30 minutes and then cooled to room temperature. This process was carefully repeated to conform five, six, and seven-bilayer photonic crystals. 

\threesubsection{2D-Perovskite}\\
A precursor solution with a concentration of 3 M was prepared by dissolving phenylethylammonium iodide ((PEA)I, Sigma Aldrich  98\%) and lead(II) iodide (PbI$_2$, Sigma Aldrich 99.999\%) in N,N-dimethylformamide (DMF, Sigma Aldrich 99.8\%) in a molar ratio of 2:1 ((PEA)I:PbI$_2$), corresponding to the stoichiometry of (PEA)$_2$PbI$_4$. The solution was stirred at room temperature for 20 minutes at 550 rpm. To promote the formation of more homogeneous thin films, additional solutions were prepared at lower concentrations of 1.3 M and 0.23 M.
Glass substrates (10 × 10 mm) were cleaned by sequential sonication in ethanol, deionised water, and ethanol again, each step lasting 10 minutes. The substrates were then dried in an oven to eliminate residual solvents. Prior to deposition, the substrates underwent plasma cleaning for 10 minutes using a Harrick plasma cleaner with air as the plasma source.
\\\\
Thin films of (PEA)$_2$PbI$_4$ were deposited by spin-coating. The deposition protocol consisted of three steps: 500 rpm for 120 s, 1500 rpm for 180 s, and 2500 rpm for 60 s, using a 0.45 $\mu$m PTFE syringe filter to dispense 0.5 mL of the solution. However, improved film quality was obtained when using a single-step spin at 5000 rpm for 30 s. Finally, the films were annealed at 100°C for 10 minutes to promote crystallization.
\\\\
\threesubsection{Microcavity}\\
The coupling of (PEA)$_2$PbI$_4$ perovskite with Bragg mirrors was carried out via spin-coating. A 1.0 cm-wide section of the photonic crystal was cut for this process. To improve film quality and surface uniformity, an intermediate PMMA layer was first deposited onto the Bragg mirror. The PMMA solution was prepared by dissolving 30 mg of PMMA in 1.0 mL of toluene.
\\\\
For the fabrication of the hybrid microcavity, the bottom Bragg mirror was first coated with a thin layer of polymethyl methacrylate (PMMA) to prevent direct electronic interaction between the perovskite and the mirror. A volume of 80\,\textmu L of PMMA solution was dispensed onto the Bragg mirror and spin-coated in two steps: 10 seconds at 100~rpm with an acceleration of 1000~rpm\,s$^{-1}$, followed by 10 seconds at 6000~rpm with an acceleration of 6000~rpm\,s$^{-1}$. The sample was then thermally annealed at 60~$^\circ$C for 5 minutes.
Next, a (PEA)$_2$PbI$_4$ perovskite film was deposited using a 0.23~M precursor solution prepared by dissolving phenylethylammonium iodide (PEAI) and lead(II) iodide (PbI$_2$) in N,N-dimethylformamide (DMF) in a 2:1 molar ratio. The solution was freshly prepared, filtered through a 0.45\,\textmu m PTFE syringe filter, and promptly used. Two drops were dispensed onto the PMMA-coated substrate and spin-coated at 5000~rpm (with an acceleration of 5000~ rpm\ s$^{-1}$) for 30 seconds. The film was then annealed at 100~$^\circ$C for 10 minutes to induce crystallization and improve structural integrity.
Finally, the perovskite layer was encapsulated with a second PMMA layer using the same spin-coating and thermal treatment protocol. This encapsulation serves both to protect the active layer from degradation and to enhance surface uniformity, thereby improving the optical quality and stability of the complete cavity. To complete the microcavity structure, a silver mirror approximately 30~nm thick was deposited on top by sputtering.

\threesubsection{Characterization}\\
The morphology and thickness of the samples were characterized using scanning electron microscopy (SEM), in a JSM 5600 LV scanning electron microscope, operating at 20 kV with a magnification of 50x.
The optical properties of all the prepared samples were analyzed by UV–Vis diffuse reflectance spectra (DRS) with a Thermo Scientific Evolution 201 spectrophotometer equipped for films measurements, in the wavelength range of 300 nm to 900 nm.  A Marienfeld Superior™ microscope glass slide was used as a reference blank. Ellipsometry measurements were performed in an Alpha 2.0 spectrometric ellipsometer (Woollam), and analyzed with CompleteEASE software; a Cauchy model was used to interpret the real part of the refractive index.
X-ray diffraction (XRD) was used to determine the crystalline structure of the (PEA)$_2$PbI$_4$ perovskite in a Bruker D-8 diffractometer with CuK$\alpha $ radiation, a Ni-0.5'\%Cu-K$\alpha $ filter in the secondary beam, and a one-dimensional position-sensitive silicon strip detector (Bruker, Linxeye). The diffraction patterns were obtained in the range of 10 ° to 40 °, with a step of 2 $\theta$ of 0.039 ° and 134.4 s per point.

\medskip
\textbf{Supporting Information} \par 
Supporting Information is available from the Wiley Online Library or from the author.

\medskip
\textbf{Acknowledgements} \par 
A.  Morales for XRD diffraction patterns, S. Tehuacanero-Cuapa and María Claudia Marchi for SEM images. L. Garrido-García for the AFM mesurments. and D. Ley-Domínguez for the silver mirrors. A.C-G. acknowledges financial support from UNAM DGAPA PAPIIT Grant No. IA101325, Project CONAHCYT No. CBF2023-2024-1765 and PIIF25. C.L.O-R. acknowledges financial support from UNAM DGAPA PAPIIT Grant IG101424.  H.A.L-G. acknowledges financial support from UNAM DGAPA PAPIIT Grant IN106725. G.P. acknowledges financial support from UNAM DGAPA PAPIIT Grant No. IN104325, Projects CONAHCYT  Nos. 1564464 and 1098652 and PIIF25. M.M-G acknowledges  to Consejo Nacional de Investigaciones Científicas y Técnicas (CONICET) for the postdoctoral scholarship with number 157248. G.JAASI acknowledges funding from Agencia I+D+i (NANOQUIMISENS, CONVE-2023-100389751-APN-MCT), CONICET (PIP-GI-11220210100917CO), and AFOSR (AWARD NO. FA9550-24-1-0209). Thanks are due to Dr. L. Malfatti and P. Rasù for their assistance in ellipsometry measurements. 

\medskip

%
%
\bibliographystyle{MSP}
\bibliography{bib}

\end{document}